\documentclass[sigconf,authorversion,nonacm,table]{acmart}
\AtBeginDocument{%
  }

\usepackage{xcolor}
\usepackage{soul}

\colorlet{usercolorname}{yellow!0}
\sethlcolor{usercolorname}

\usepackage{graphicx}
\usepackage{subcaption}

\usepackage{multirow}

\hypersetup{
  pdftitle={Access Over Deception},
  pdfauthor={Tobias Pellkvist, Katie Seaborn, Miu Kojima},
  pdfsubject={Fighting Deceptive Patterns through Accessibility},
  pdfkeywords={Dark Patterns, Deceptive Design, Manipulative User Interfaces, User Interface Design, Web Accessibility, Web Content Accessibility Guidelines (WCAG), Heuristic Evaluation, Design Ethics},
  pdflang={en-CA}
}

\copyrightyear{2026}
\acmYear{2026}
\setcopyright{cc}
\setcctype{by-nc-nd}
\acmConference[CHI '26]{Proceedings of the 2026 CHI Conference on Human Factors in Computing Systems}{April 13--17, 2026}{Barcelona, Spain}
\acmBooktitle{Proceedings of the 2026 CHI Conference on Human Factors in Computing Systems (CHI '26), April 13--17, 2026, Barcelona, Spain}
\acmPrice{}
\acmDOI{10.1145/3772318.3791053}
\acmISBN{979-8-4007-2278-3/2026/04}

\usepackage{pdfx}
\begin{document}

\title[Access Over Deception]{Access Over Deception: \texorpdfstring{\\}{} Fighting Deceptive Patterns through Accessibility}

\author{Tobias Pellkvist}
\orcid{0009-0006-9256-1834}
\affiliation{%
  \institution{Institute of Science Tokyo}
  \city{Tokyo}
  \country{Japan}
}
\affiliation{%
  \institution{TU Wien}
  \city{Vienna}
  \country{Austria}
}
\email{e12024024@student.tuwien.ac.at}

\author{Katie Seaborn}
\orcid{0000-0002-7812-9096}
\affiliation{%
  \institution{Institute of Science Tokyo}
  \city{Tokyo}
  \country{Japan}
}
\affiliation{%
  \institution{University of Cambridge}
  \city{Cambridge}
  \country{UK}
}
\email{katie.seaborn@cst.cam.ac.uk}

\author{Miu Kojima}
\orcid{0009-0006-6122-6750}
\affiliation{%
  \institution{Institute of Science Tokyo}
  \city{Tokyo}
  \country{Japan}
}
\email{kojima.m.69f5@m.isct.ac.jp}

\renewcommand{\shortauthors}{Pellkvist et al.}

\begin{abstract}
  Deceptive patterns, dark patterns, and manipulative user interfaces (UI) are a widely used design strategy that manipulates users to act against their own interests in pursuit of shareholder aims. These patterns may particularly affect people with less education, visual impairments, and older adults. Yet, access is a critical feature of the user experience (UX), development standards, and law. We considered whether and how the Web Content Accessibility Guidelines (WCAG) and related legislation, like the European Accessibility Act (EAA), could act as a tool against deceptive patterns. We used heuristic evaluation to analyze whether and how deceptive patterns violate or conform to these guidelines and legal statutes. Although statistical analysis revealed no significant differences by pattern type, we identified three patterns implicated by the WCAG guidelines: Countdown Timer, Auto-Play, and Hidden Information. We offer this approach as one tool in the fight against UI-based deception and in support of inclusive design.
\end{abstract}

\ccsdesc[500]{Social and professional topics~Codes of ethics}
\ccsdesc[500]{General and reference~Computing standards, RFCs and guidelines}
\ccsdesc[300]{Human-centered computing~Heuristic evaluations}
\ccsdesc[500]{Human-centered computing~Accessibility design and evaluation methods}
\ccsdesc[100]{Human-centered computing~Accessibility systems and tools}
\ccsdesc[300]{Human-centered computing~Web-based interaction}
\ccsdesc[300]{Software and its engineering~Graphical user interface languages}
\ccsdesc[300]{Human-centered computing~Interaction design theory, concepts and paradigms}

\keywords{Dark Patterns, Deceptive Design, Manipulative User Interfaces, User Interface Design, Web Accessibility, Web Content Accessibility Guidelines (WCAG), Heuristic Evaluation, Design Ethics}


\maketitle

\section{Introduction}
\label{sec:intro}
Deceptive patterns---also called dark patterns, manipulative user interfaces (UI), and deceptive designs, and hereafter ``DPs''---are elements of the UI that trick users or prevent user action in a way that benefits the shareholders and often disadvantages the user~\cite{Gray2024,Mathur2021}.
DPs continue to be an unfortunate and prevalent part of online life, most notably in shopping websites~\cite{Mathur2019} and social media~\cite{Mildner2023DarkArts}. DPs work by exploiting cognitive biases~\cite{Mathur2019}, manipulating and restricting choices~\cite{Susser2019Technology, Mathur2021}, and leveraging social engineering~\cite{Fogg2002}, among other strategies, leading to cognitive load~\cite{Mathur2021}, reduced user agency~\cite{Ye2025}, negative emotions~\cite{Mathur2021}, and unintended spending~\cite{Mathur2019}. While affecting all users, certain types of DPs can negatively impact specific user groups, like people with less education~\cite{Luguri2021, BongardBlanchy2021}, children~\cite{schaeffer2024},  older adults~\cite{aarp2011report, BongardBlanchy2021, sanchez2024manipulative}, and those with disabilities, notably visual impairments~\cite{Kodandaram2023, Lewis2025}. Only recently has the intersection of DPs and accessibility been considered. 
People with disabilities encounter unique and greater harms when subjected to DPs~\cite{Kodandaram2023, Lewis2025}. This includes more financial costs, like accidental purchases~\cite{Kodandaram2023} and unexpected fees and subscriptions, cognitive costs through disorienting designs, like countdown timers with constant alerts, and time costs~\cite{Lewis2025}. In short, people with disabilities may suffer more and differently when encountering interface-based deception.

While efforts to combat DPs exist, including  awareness campaigns like \href{https://www.deceptive.design}{deceptive.design} and lawsuits like the FTC vs. Epic Games (\autoref{sec:combatdps}), a surefire solution remains elusive. For one, defining what DPs are 
is difficult, with expert consensus hard to achieve~\cite{Mathur2021,Gray2024}. Some definitions, for example, include designer intent, even though this can be difficult to judge~\cite{Gray2018,Conti2010}. Subsequently, actionable legislation has been hard to establish~\cite{Yi2024,Brenncke2023}. 
For example, the European Digital Services Act (DSA)~\cite{dsa} has attempted to define and dispense prohibitions against DPs. Yet, concerns have been raised about its effectiveness due to use of ambiguous terminology~\cite{Mathur2021,Gray2024} and loose definitions~\cite{Yi2024,Brenncke2023}. We note, however, that its novelty (enacted in 2024) prevents a full understanding of its true effectiveness.
Concerns also relate to overregulation and limiting designers unfairly~\cite{Yi2024}. Furthermore, difficulties have arisen when justifying legal cases where the harm was minor or the company was too small to warrant expensive lawsuits with uncertain outcomes~\cite{Luguri2021}. This has left the community searching for methods and tools to identify and combat DPs.

One strategy is to leverage existing guidelines, standards, and regulations that intersect with DPs~\cite{Yi2024}. We propose that accessibility legislation and standards may be an effective and inclusive option. 
While some research has considered DPs against accessibility, this work has been qualitative and case study-focused~\cite{Kodandaram2023, Lewis2025}, leaving the general relationship between DPs and established standards unknown. A clear starting point is the Web Content Accessibility Guidelines (WCAG), created by the international standards body \href{https://www.w3.org/}{World Wide Web Consortium (W3C)}. The WCAG have recently become a critical part of many regulations and standards (\autoref{sec:wcag}). In principle, DPs that are inaccessibly designed should not be legally allowed or endorsed in practice, as per the WCAG against DPs. However, no work has explored or offered proof-of-concept evidence of DPs violating accessibility standards.

An accessibility standards approach provides unique benefits over other methods. The WCAG offers a set of strictly defined guidelines---and a certain level of objectivity---even while some guidelines leave room for interpretation and design creativity. Accessibility standards could also complement and be usefully integrated with legislation, existing and in-progress. For instance, the Digital Fairness Act (DFA)~\cite{dfa}, an upcoming extension of the DSA targeting DPs and similar practices, notes the gaps for certain forms of interface deception that are also linked to accessibility issues (\autoref{sec:combatdps}). Additionally, the existing, robust set of WCAG tools (\autoref{sec:wcag} and \autoref{sec:discuss}) could be used in combination with DP detection methods to quantitatively identify overlapping sets of DPs and accessibility violations. Since achieving general accessibility is becoming more important across societies worldwide,  responsible designers and companies aiming for inclusion may also be incentivized to decrease reliance on DPs or avoid them altogether.
 
To establish the efficacy of using accessibility standards to fight DP use, we aimed to unearth the relationship between DPs and the WCAG guidelines. We asked: \textbf{How do different types of DPs fare by accessibility level according to the WCAG?} For this, we carried out a heuristic evaluation~\cite{Nielsen1990} of selected DP examples, identifying which DPs violated what guidelines. As \emph{WCAG Version 2.1, Level AA} covers the most relevant standards and laws, we considered this combination of version and compliance level as a minimum accessibility requirement for DP examples to fulfill. 
We combined manual analysis and multiple WCAG evaluation tools. 
Our premise was: If an alternative design that preserves both function and accessibility was not possible, the pattern must violate the WCAG standard and by extension the laws mentioned above. In short, the WCAG may be used to halt use of these DPs.

We contribute:
\begin{itemize}
    \item The first proof-of-concept use of accessibility standards to combat DPs. We reveal the efficacy for specific DP types, establishing the relationship between DPs and the WCAG by identifying violated guidelines per DP example. Thus, DPs generally likely to be inaccessible were identified. 
    \item Leveraging the ontology by \citet{Gray2024}, identification of three DP types 
    implicated by the WCAG guidelines: Countdown Timers, Auto-Play, and Hidden Information. 
\end{itemize}

\section{Background}
\label{sec:bg}

\subsection{Web Accessibility and Web Content Accessibility Guidelines (WCAG)}
\label{sec:wcag}
There are many ways to evaluate the website accessibility. User testing and expert evaluation can assess accessibility in a more user-centred and qualitative way. Alternatively, the \href{https://www.w3.org/WAI/standards-guidelines/wcag}{WCAG}~\cite{WCAG1,WCAG2}, a set of guidelines (currently version 2.2~\cite{WCAG2}) authored by the \href{https://www.w3.org/}{W3C}, are primarily used by developers to make websites more accessible~\cite{WCAG1,WCAG2}. WCAG success criteria are grouped into three conformance levels (A, AA, AAA), with higher levels including all lower-level criteria. The AAA level represents the highest conformance to the WCAG. Some guidelines need manual assessment, but a subset can be evaluated by automated tools, although coverage differs by tool~\cite{Lempola2024,Pool2023}.  

While the WCAG guidelines are a prominent tool for web accessibility~\cite{Filipe2023}, they are not flawless. For example, studies have shown that the guidelines only cover $\sim$50\% of manually found accessibility issues~\cite{Power2012}, as well as have low compliance rates, especially when analyzing cookie banners~\cite{Clarke2024}. Developers can experience difficulties with understanding the guidelines~\cite{Bubich2024}. Others have found a lack of sufficient coverage of issues that affect people with non-visual impairments like cognitive disabilities~\cite{Gartland2022}.
Comparing the usage of WCAG tools to user testing in combination with expert evaluation has shown that the latter is more effective at finding more issues~\cite{Mateus2021}. 

However, there is value in using automated tools in combination with manual evaluation. This approach is more cost-effective than user testing and expert evaluation, and might reveal issues that are usually not discovered by qualitative approaches~\cite{Mateus2021}. The accuracy of automated tools may improve with further development. The W3C continually improves the WCAG; currently, they are working on \href{https://www.w3.org/WAI/standards-guidelines/wcag/wcag3-intro/}{WCAG 3.0}, aiming to make the guidelines ``easier to understand'' and ``cover more user needs, including more needs of people with cognitive disabilities''~\cite[\S WCAG 3 Draft approach]{WCAG3}.
This addresses some of the issues identified by previous research~\cite{Bubich2024, Power2012, Gartland2022}. The WCAG has also recently been included as part or as a basis for laws and standards, like the \href{https://commission.europa.eu/strategy-and-policy/policies/justice-and-fundamental-rights/disability/union-equality-strategy-rights-persons-disabilities-2021-2030/european-accessibility-act_en}{European Accessibility Act (WCAG 2.1 AA)}, \href{https://waic.jp/docs/jis2016/understanding/201604/}{JIS X 8341-3 (Japanese Standard - WCAG 2.0)}, and \href{https://www.iso.org/standard/58625.html}{ISO/IEC 40500:2012 (WCAG 2.0)}, solidifying its legitimacy. However, we note that the WCAG is itself not legislation; its efficacy relates to baseline accessibility compliance when formalized in the law. Lawmakers may extend what a given legislation covers, but the WCAG provides a solid foundation.

\subsection{Deceptive Patterns and Design Strategies}
\label{sec:dp}
Deceptive patterns are design strategies implemented into UI that lead users to make decisions in favour of shareholders, companies, or services utilizing them, often acting against the user's own interests. This is done by exploiting cognitive biases~\cite{Mathur2019}, manipulating and restricting user choice, and otherwise intentionally inconveniencing users~\cite{Susser2019Technology, Mathur2021}. 
This causes a range of negative effects on users, like cognitive (over)load~\cite{Mathur2021}, reduced agency~\cite{Ye2025}, negative emotions, like frustration and anger~\cite{Mathur2021}, and unwanted and unneeded expending of time, money, and other resources~\cite{Mathur2019}.
They can, for  instance, capture user attention~\cite{Narayanan2020,Ye2025}, especially on social media and SNS, in pursuit of spending more time on the platform. While a concrete, agreed-upon operationalization remains elusive~\cite{chang_theorizing_2024,Mathur2021,Gray2024}, the guiding idea is that DPs prioritize the deployer over the user, regardless of user awareness and despite potential or actual user harm.

\subsubsection{Additional Effects Relevant to Accessibility}
The HCI community has only recently begun to scratch the surface of the intersection between DPs and accessibility. Two studies, focusing on visual impairments, have shown that people using screen readers fall prey more often to Disguised Ads~\cite{Kodandaram2023} and experience more intense negative effects through higher costs, decreased agency, and unique issues, like distracting voiced Countdown Timers creating higher cognitive load~\cite{Lewis2025}. These issues have led some users to avoid using services where DPs are prevalent~\cite{Lewis2025}, rendering DPs a matter of basic access. This was named an ``anti-pattern'' by one participant. 

Certain user groups and people with certain characteristics may also be affected to a greater degree, including people with less education failing to recognize and falling prey more often to DPs~\cite{Luguri2021, BongardBlanchy2021}. Reports by the FTC (Federal Trade Commission) and the AARP (American Association of Retired Persons) suggest that people with lower incomes might fall for DPs more due to income-implicated factors like smaller screens on older devices, and that older adults fall for fraud more often~\cite{ftc2025darkpatterns, aarp2011report}. Even though fraud and DPs are not strictly the same, \citet{BongardBlanchy2021} have shown that older people fail to recognize DPs as often as younger people, but fall for DPs more often.

\subsubsection{Deceptive Pattern Ontology}
\label{sec:ontology}
Given no clear and consistent definition of DPs and DP types~\cite{Mathur2021, Gray2024}, we used the ontology proposed by~\citet{Gray2024} in 2024. This recent and comprehensive ontology, vetted by the CHI community, is split up into three categories, High-, Meso- and Low-level, where each Low-level type belongs to a Meso-level type, which in turn belongs to a High-level type. The subcategory approach classifies and sorts DPs into a hierarchical structure. In total, there are five High-level types and 43 Low-level patterns. However, some Meso-level patterns do not have associated Low-level patterns. For ease of reading and internal consistency, we refer to all Meso-level patterns without Low-level patterns and Low-level patterns as ``lowest-level'' in this paper.

\subsection{Using Accessibility to Combat DPs}
\label{sec:combatdps}
Arguing against the use of DPs has proven tricky due to the noted ambiguities and ubiquity. Companies have been held legally accountable; for example, the FTC's successful lawsuit against Epic Games due to unexpected payments in ``Fortnite''~\cite{ftc2024fortnite}. Existing and upcoming acts, legislation, and regulatory standards offer methods but may not yet actionable. As noted, the DSA~\cite{dsa} and other EU legislation include and define DPs, but also bear unclear definitions and phrasing, such that regulators struggle when judging cases~\cite{Yi2024, Brenncke2023}. 
Autonomy, for instance, is a common keyword, but is not well-defined, creating barriers to its use as a guiding ``normative lens'' of what ought to be when making judgments about DPs~\cite{Brenncke2023,Ahuja2022autonomy}. This has generated efforts to define autonomy and map out its dimensions within the context of interface deception~\cite{Ahuja2022autonomy,santos2024online,Brenncke2023}. 
\citet{Ahuja2022autonomy} analyzed 151 DPs against 16 taxonomies to derive four dimensions of autonomy: agency, freedom of choice, control, and independence.
\citet{Ahuja2025autonomy} targeted the three DSA-prohibited types of autonomy violation~\cite{dsa}: deception, manipulation, and distortion/impairment. They showed how 59 DPs intersected with these types in an 8-part framework divided between information and choice spaces.
This scholarship has yet to be adopted into law or actionable directives, but provides a case study for heuristic arguments.
Due to these complications, other approaches have been proposed, like private law (i.e., cases between private parties, e.g., consumer vs. company)~\cite{Dickinson2023} and laws related to sectors outside consumer protection~\cite{Yi2024}.

The notion of a multi-pronged solution and leveraging existing regulations inspired us to consider use of web accessibility standards (\autoref{sec:wcag}). 
Adherence to accessibility standards could prevent the use of DPs that rely on inaccessible designs. 
Legal guidelines may be bolstered in combination with such standards, which could specify noncompliant designs based on vetted principles (via the W3C) and a suite of existing detection tools. The upcoming DFA~\cite{dfa}, for instance, could be fortified by use of the WCAG to justify how ``giving more prominence to certain choices'' (Article 25, 3a) is violated. Misusing color contrast or text size to make certain options prominent over others would fundamentally violate accessibility guidelines for visually-impaired people.
We advocate for accessibility standards as a short-term substitute for and long-term complement to the weaknesses in consumer protection legislation. However, as noted in \autoref{sec:discuss} and \autoref{sec:limits}, there may be inherent usability advantages---effectively, DP countermeasures---and disadvantages to WCAG compliance. Here, as a first step, we attempted to demonstrate the feasibility and utility of leveraging accessibility legislation against interface deception.

\section{Methods}
\label{sec:methods}

We conducted a heuristic evaluation~\cite{Nielsen1990} of real web-based DPs, combining manual coding and automatic accessibility assessment tools. Our protocol was registered before data analysis on OSF\footnote{\url{https://osf.io/fsdph}}.

\subsection{Procedure}
\label{sec:procedure}
Real-world examples were gathered (\autoref{sec:sampling}), categorized by type based on the \citet{Gray2024} ontology, and then evaluated with the WCAG. For this, we mapped the DPs to the lowest-level DP categories (defined in \autoref{sec:ontology}). This way, we avoided over-generalizing, achieved precision in identifying DP types (which are linked to specific mechanisms), and enabled coding consistency among raters while supporting future replicability efforts. 

The evaluation was carried out in two ways. 

First, we used three tools---\href{https://wave.webaim.org/}{WAVE}, \href{https://www.ibm.com/able/toolkit/}{IBM Equal Access Accessibility Checker}, and \href{http://qualweb.di.fc.ul.pt}{QualWeb} 
listed officially by the \href{https://www.w3.org/WAI/test-evaluate/tools/list/}{W3C Accessibility Initiative (WAI)}. As browser extensions, they made isolation of violated guidelines relevant to DPs easier to detect \emph{in situ}. They are complimentary, achieving higher coverage of different types of accessibility problems and high instance counts, i.e., higher probability of finding problems when used together~\cite{Pool2023, Lempola2024}. WAVE and QualWeb, for example, allow for a high increase in issue counts found when used together~\cite{Pool2023}, while WAVE and IBM cover different types of accessibility problems~\cite{Lempola2024}. Additionally, while rarely needed, we used the \href{https://webaim.org/resources/contrastchecker/}{WebAIM color contrast tool} for screenshots when the deception was based on visuals. 
The Google Chrome versions of the extensions were used.

The second was manual analysis for when the tools did not evaluate all guidelines~\cite{Mateus2021}. The lead researcher checked for WCAG violations manually. Another researcher also checked a random 30\% sample ($n=20$). Inter-rater reliability (IRR) was assessed with Cohen's kappa at a minimum $\kappa$=.75~\cite{Cohen1960} for both DP types and accessibility (issue present vs. not present). When raters disagreed, the case was re-evaluated through discussion. 
Note that accessibility problems found through tools and manual analysis are not later differentiated, since baseline results on access was the goal. 

Four of 20 examples were disqualified due to unreproducible behaviour, like missing countdown timers, invalid links, or unclear instructions.
For the remaining 16, we calculated $\kappa$ for two different ratings: agreement on accessibility per example and by type. 
Since we used two predefined sets of rules (the ontology~\cite{Gray2024} and the WCAG), we consolidated DP types and missed accessibility issues and re-examined the definitions when agreement could not be found. A notable factor was the element composition of the DP. 
We struggled to make judgments about whether, for example, the alternative text describing a picture of a person was relevant to an example of Endorsements and Testimonials. We achieved a $\kappa$=.81, $\kappa$=.75, and $\kappa$=.78 for accessibility, DPs, and total mean, respectively.

\subsection{Sampling}
\label{sec:sampling}

The data (web-based examples of DPs) was drawn from the real world, using collections like halls of shame\footnote{\url{https://deceptive.design/hall-of-shame}}\footnote{\url{https://hallofshame.design/collection}} and community submissions on the \href{https://darkpatternstipline.org}{Dark Patterns Tip Line} and \href{https://reddit.com/r/assholedesign}{Reddit}.
We also used \href{https://chatgpt.com}{ChatGPT} and \href{https://perplexity.ai}{Perplexity} prompts like ``Where can I find live examples for these dark patterns?'' and ``Give me collections of dark pattern examples,'' leading to collections like the one by \citet{knownhost} and specific websites like \href{https://booking.com}{Booking.com} and \href{https://jetstar.com/jp/en}{Jetstar}.

Examples of DPs were often fixed. 
Where necessary, the \href{https://web.archive.org/}{Wayback Machine}, the search engine for the \href{https://archive.org/}{Internet Archive}, an archive of all public websites saved by web crawlers, was used to confirm the website described in the example/s. If inaccessible through the Internet Archive, we searched for new examples.

We attempted to find three+ cases for each of the 43 lowest-level patterns~\cite{Gray2024}, aiming for 129+ cases and a consistent minimum.
Given time constraints, the nature of most DP collections (screenshots instead of websites, which cannot be used in WCAG evaluations), and low variety in DPs from the sources used, we ultimately found and analyzed 68 DPs across 26/43 different types, with varying example counts.

\subsection{Variables}
\label{sec:variables}
We used two metrics: counts of accessible examples (no issues present) and inaccessible examples (1+ issue present). These were recorded in a matrix by DP type and analyzed with Fisher's exact tests (\autoref{sec:analysis}). Given the high variance in examples by DP, we also compared the normalized counts $n$ for each DP using Chi-squared tests.
\begin{displaymath}
n(DP) = \frac{\sum_{}accessibility\;issues}{example\;count}
\end{displaymath}
 Analyzing both metrics allowed for comparison between the numbers of issues and accessible examples.
We did not use a common metric for WCAG analysis---the compliance rate (ratio of guidelines followed vis-\`a-vis all guidelines)---because we only analyzed parts of whole websites, which would inflate the average resulting ratio.

\subsection{Analysis}
\label{sec:analysis}
We used the Chi-squared and Fisher's exact tests. The Chi-squared test was used for the normalized count of accessibility issues. 
Fisher's tests were applied to the counts of accessible vs. inaccessible examples when counts were low. Tests were applied to compare the lowest-level patterns (43 patterns) and High-level patterns (Forced Action, Interface Interference, Obstruction, Sneaking, Social Engineering). We used the standard $p<.05$ criteria for significance.

\section{Results}
\label{sec:results}
\begin{figure*}[ht]
  \centering
  \includegraphics[width=\textwidth]{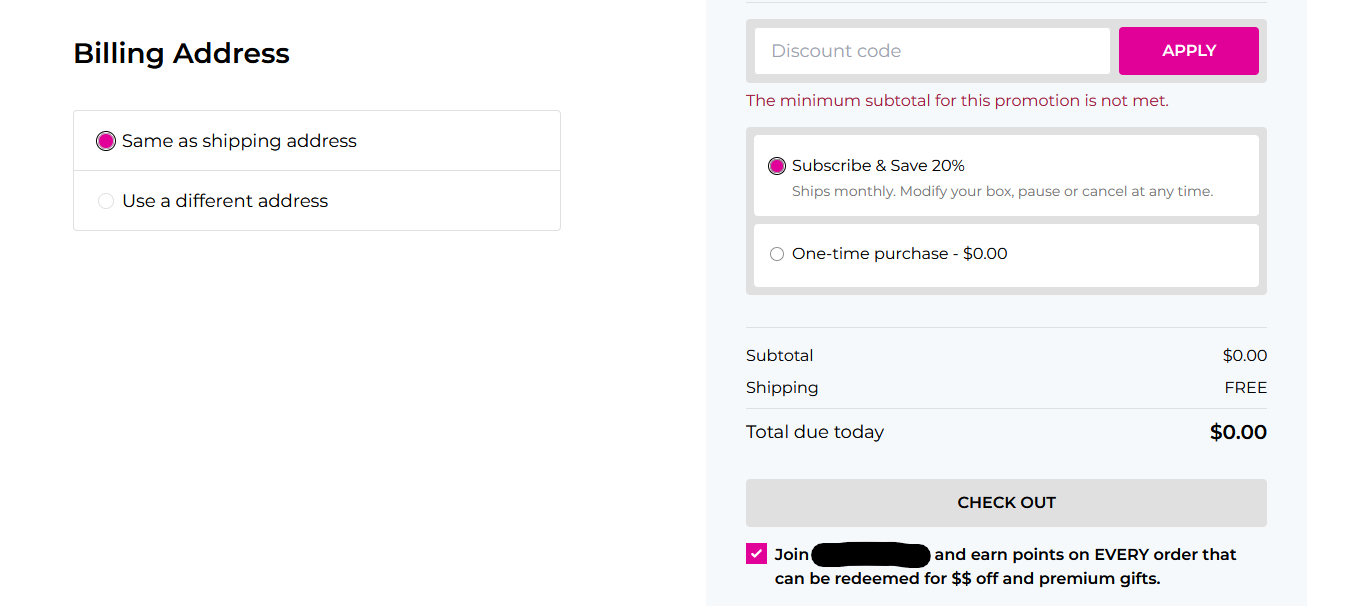}
  \caption{Example of Bad Defaults. The ``Subscribe \& Save 20\%'' and ``Join program'' options are preselected. The WCAG tools revealed insufficient contrast between the text and background in places, which could limit an informed choice among options.}
  \Description{Image showing an example of Bad Defaults. The ``Subscribe \& Save 20\%'' radio input and ``Join program'' checkbox are selected by default. The results of the analysis illustrated here show contrast issues between text and background, possibly limiting informed choices.}
  \label{fig:dp-ex}
\end{figure*}

\begin{figure}[ht]
  \centering
  \includegraphics[width=.4\textwidth]{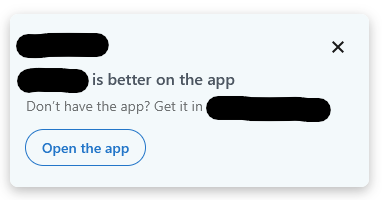}
  \caption{Example of Nagging. A pop-up about a desktop app being available reappears on different pages, even when dismissed. The WCAG tools revealed insufficient contrast between the gray text and background, potentially obscuring the reason for the pop-up.}
  \Description{Image showing an example of Nagging. A pop-up reminds the user that a desktop app is available, which reappears on different pages, when dismissed. The results of the analysis illustrated here show contrast issues between text and background, possibly obscuring the reason for the pop-up.}
  \label{fig:dp-ex2}
\end{figure}

Across all 26 types, we found 68 examples with a total of 105 WCAG guideline violations. An overview is presented in \autoref{tab:all-data} for all high-level DPs. Examples 
are in \autoref{fig:dp-ex} and \autoref{fig:dp-ex2}.

\begin{figure}[ht]
  \centering
  \includegraphics[width=.34\textwidth]{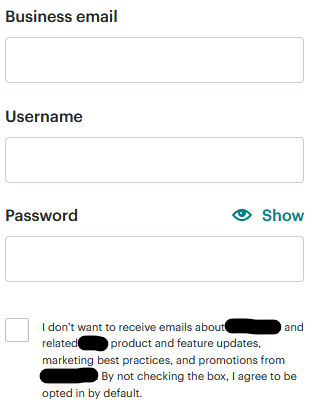}
  \caption{Example of Trick Questions. Clicking the checkbox will opt-out of the subscription, even though this is usually a opt-in checkbox. No accessibility issues were detected.}
  \Description{Image showing an example of Trick Questions. The image shows a registration form, clicking a checkbox with confusing text will opt-out of a mail subscription, even though this is usually a opt-in checkbox. No accessibility issues were detected.}
  \label{fig:dp-ex3}
\end{figure}

\begin{table*}
  \caption{Counts and normalized counts for each high-level DP.}
  \label{tab:all-data}
  \begin{tabular}{l|ccccc}
    \toprule
     & Forced Action & Interface Interference& Obstruction& Sneaking& Social Engineering \\
    \midrule
    Counts & 17& 51& 3& 22& 12 \\
    Normalized Counts & 1.33& 1.75& 0.75& 2.1& 0.71\\
  \bottomrule
  \end{tabular}
\end{table*}

\subsection{Prevalence of Accessibility Issues}
Given the small counts and the use of normalized data, the Chi-squared tests should be interpreted with caution.

\subsubsection{High-Level DPs}
\begin{figure*}[ht]
  \centering
  \includegraphics[width=.575\textwidth]{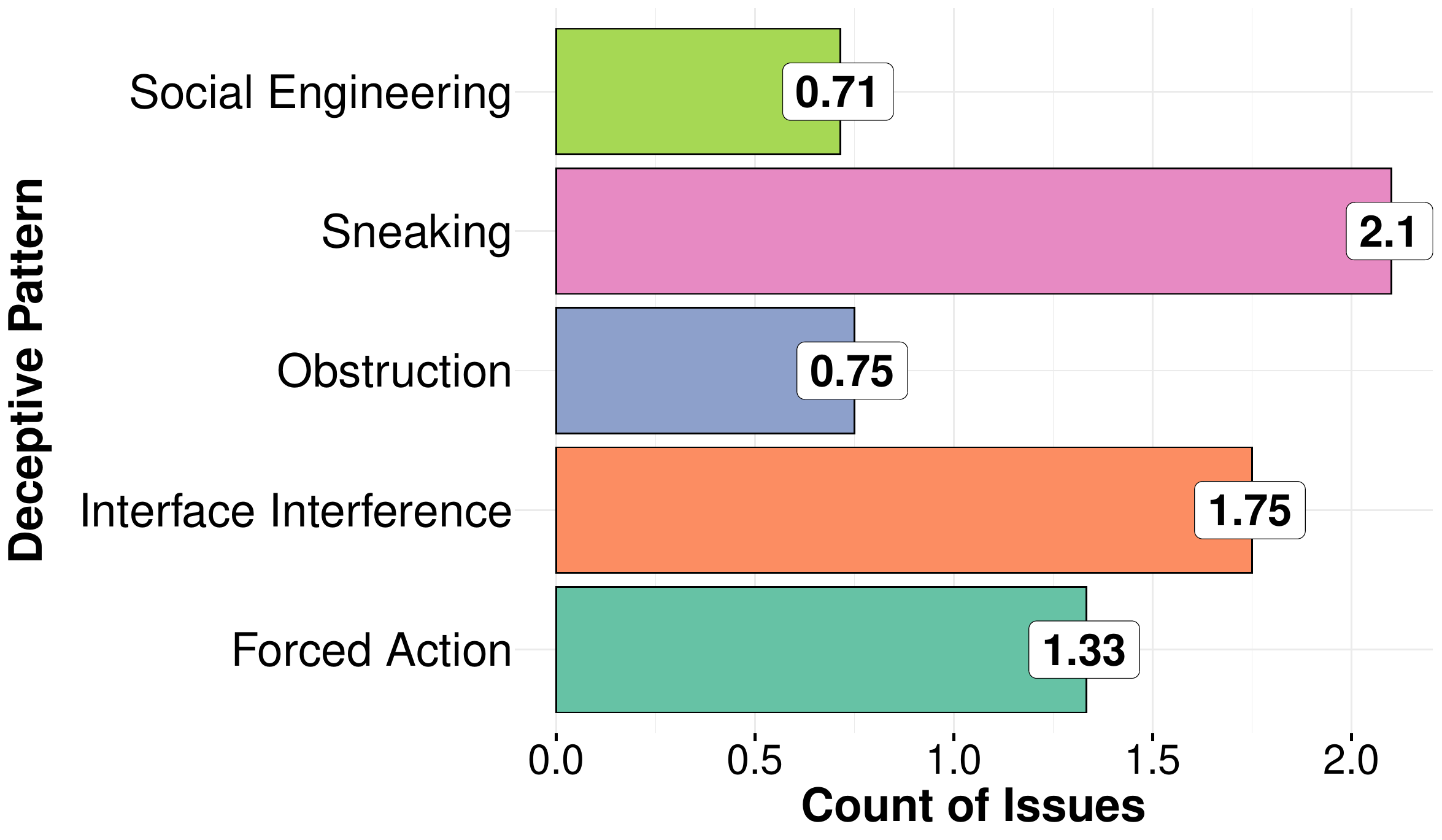}
  \caption{Bar chart showing the distribution of normalized accessibility issue counts per DP type for high-level DPs.}
  \Description{A horizontal bar chart showing the distribution of normalized accessibility issue counts (on the x-axis) per DP type for all 5 high-level DPs (located on the y-axis): Forced Action, Interface Interference, Obstruction, Sneaking and Social Engineering. Sneaking is highest, while Obstruction and Social Engineering are lowest.}
  \label{fig:main-counts}
\end{figure*}

Prevalence varied by DP type.
The Chi-squared test for the high-level data did not find evidence that the normalized counts are not uniformly distributed across high-level DP types 
($\chi^{2}(4,N=6.65)=1.12,p=.89)$. Notably, Obstruction and Social Engineering had the lowest counts, between 0.7 to 0.75 (\autoref{fig:main-counts}). Sneaking had the highest at 2.1, followed by Interface Interference with 1.75, then Forced Action with a count of 1.33.

\subsubsection{Lowest-Level DPs}
The Chi-squared test for the lowest-level data, as for the high-level, did not find evidence that the normalized counts are not uniformly distributed across the lowest-level DP types 
($\chi^{2}(25,N=34.33)=28.92,p=.26)$. Notable DPs include Sneak Into Basket and Privacy Zuckering (more than 3 issues) and Confirmshaming, Dead End, Low Stock, etc. (less than .5 issues).

\begin{figure*}[ht]
  \centering
  \includegraphics[width=\textwidth]{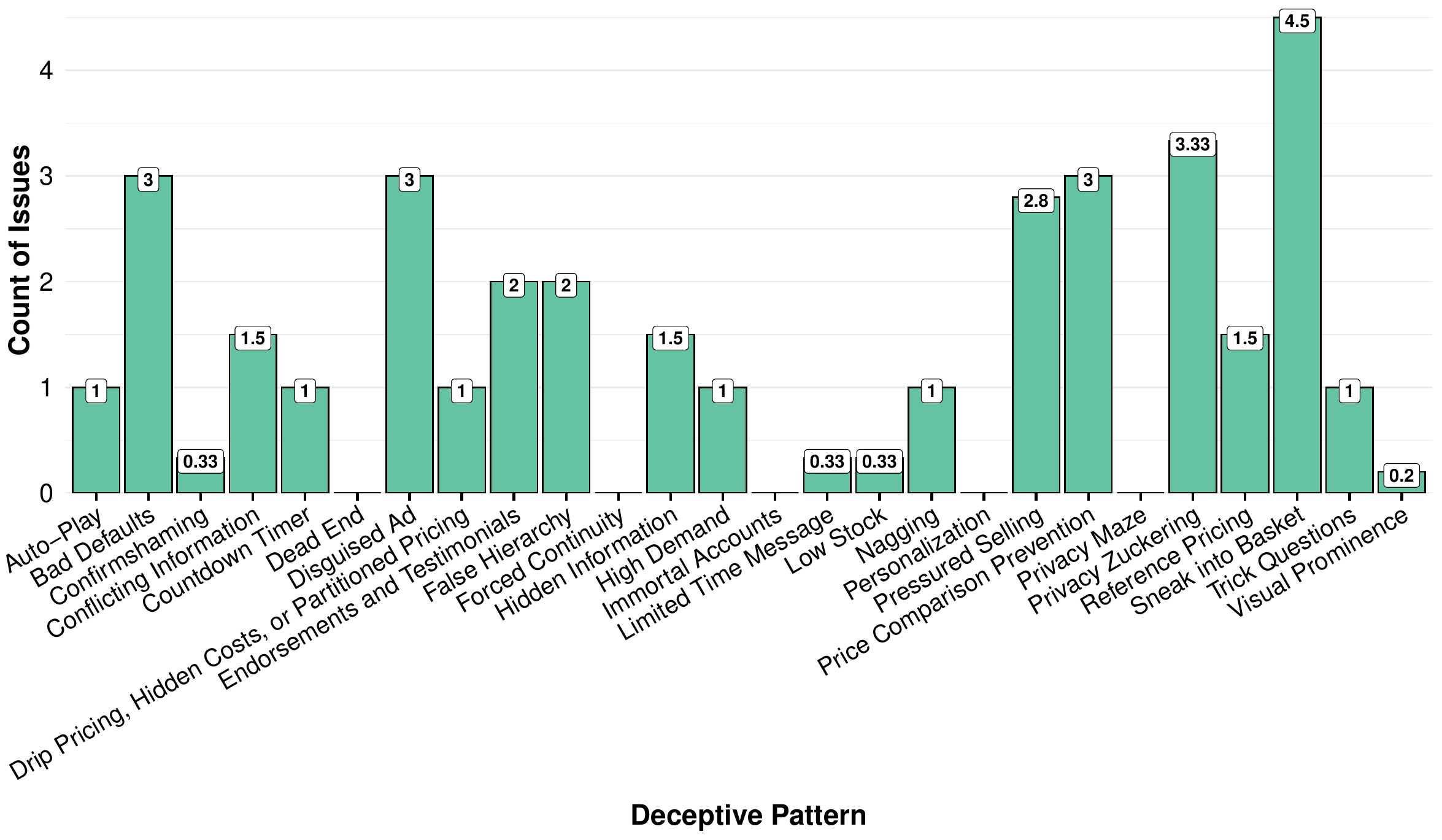}    
  \caption{Bar chart showing the distribution of normalized accessibility issue counts per DP type for lowest-level DPs.}
  \Description{A vertical bar chart showing the distribution of normalized accessibility issue counts per DP type for lowest-level DPs. Sneak Into Basket is the highest with 4.5, while Confirmshaming, Dead End, Forced Continuity, Immortal Accounts, Limited Time Message, Low Stock, Personalization, Privacy Maze and Visual Prominence have either no or almost no accessibility issues (less than 0.5).}
  \label{fig:lowest-counts}
\end{figure*}

\subsection{Prevalence of Examples with Issues vs. No Issues}

\subsubsection{High-Level DPs}
\begin{figure*}[ht]
  \centering
  \includegraphics[width=.575\textwidth]{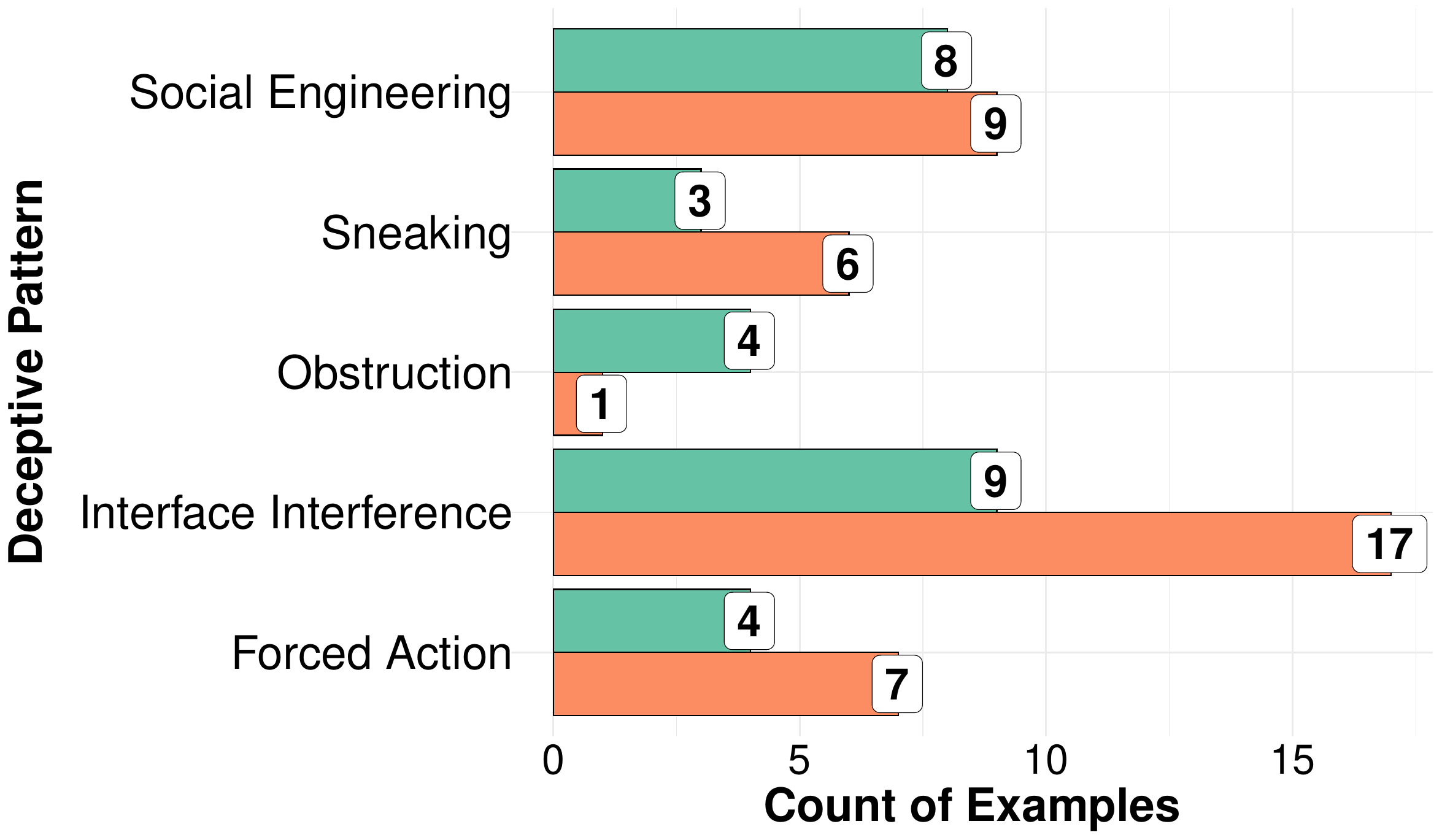}
  \caption{Bar chart showing the counts of examples that have accessibility issues (orange, lower) vs. examples without accessibility issues (green, upper) per high-level DP type.}
  \Description{A paired bar chart showing the counts of examples that have accessibility issues vs examples that have no accessibility issues per high-level type. The lower bar shows the count of examples that are inaccessible and the upper bar shows the count of examples that are accessible for each DP type. Interface Interference, Forced Action and Sneaking show an approximate 2:1 ratio, Social Engineering has slightly more inaccessible than accessible examples, while Obstruction shows a ratio of 1:4. Interface Interference has the highest total most amount of examples, while Obstruction has the lowest.}
  \label{fig:main}
\end{figure*}

A Fisher's exact test for high-level DP types indicated no relationship between DP type and number of inaccessible examples ($p =.42, \alpha =0.05$). Notably, Obstruction was the only high-level type with more accessible examples than non-accessible examples (\autoref{fig:main}).

\subsubsection{Lowest-Level DPs}
A Fisher's exact test bore no indication of a relationship between DP type and number of inaccessible examples ($p =.45, \alpha =0.05$). 
A larger sample size may be indicative. For example, Bad Defaults had 3 times and Pressured Selling 4 times as many examples with issues as examples with no issues. Conversely, Visual Prominence had substantially fewer examples with issues.

\begin{figure*}[ht]
  \centering
  \includegraphics[width=\textwidth]{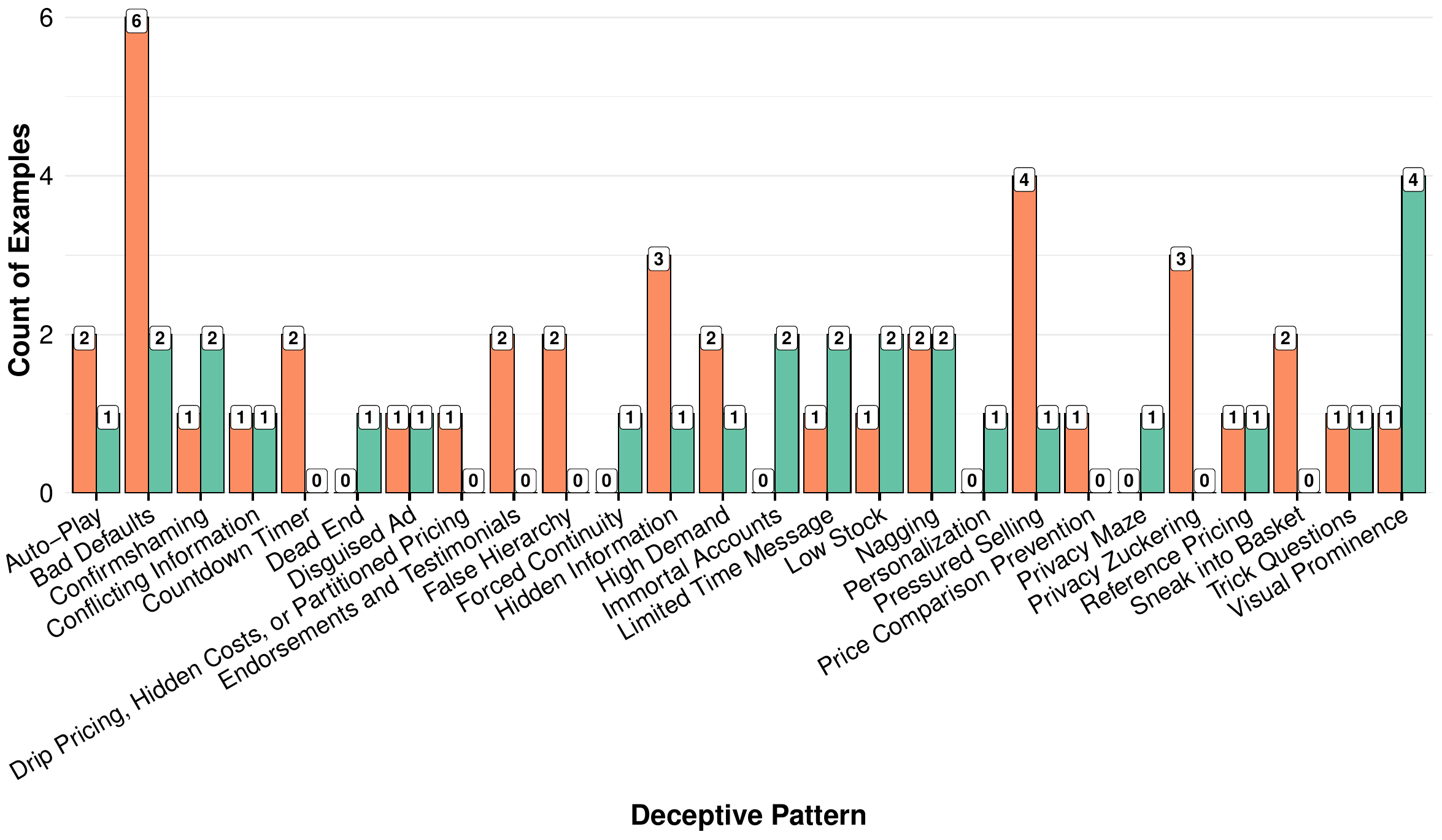}
  \caption{Bar chart showing the counts of examples with accessibility issues (orange, left) vs. examples without accessibility issues (green, right) per lowest-level DP type.}
  \Description{A paired bar chart showing the counts of examples that have accessibility issues vs examples that have no accessibility issues per lowest-level type. The left bar shows the count of examples that are inaccessible and the right bar shows the count of examples that are accessible for each DP type. Bad Defaults, High Demand and Pressured Selling stand out for having relatively high inaccessible examples, while Visual Prominence stands out for having relatively little inaccessible examples.}
  \label{fig:lowest}
\end{figure*}

\subsection{Observed Relationships by DP Type and Guideline}
A few designs tied to DP types directly contradicted 1+ WCAG guidelines and would thus require a ``fundamental change'' to become accessible: a visual or functional modification to the core properties that define the DP \emph{as} a DP, such that it would no longer be a DP. For example, since Trick Questions use confusing wording to deceive users, changing contrast would not be a fundamental change, but changing the wording to make the sentence less confusing would be---which would eliminate the deception. 

Three of 26 were identified as having an example requiring a fundamental change: Countdown Timers, Auto-Play, and Hidden Information. Still, none were recognized as inaccessible in every instance, indicating strong ties to the specific design of the DP. We next describe why each contradicts which WCAG guideline(s).

\subsubsection{Countdown Timers}
All three analyzed examples were considered inaccessible by design because they inherently contradict the guidelines \emph{2.2.1 Timing Adjustable} and \emph{2.2.2 Pause, Stop, Hide}. Guideline \emph{2.2.1} states conditions for time limits, while \emph{2.2.2} inhibits moving, blinking, scrolling, or auto-updating information.

Given these guidelines, the following limitations can be inferred:
\begin{enumerate}
    \item Turning Countdown Timers off, as well as adjusting or extending the time, is necessary (2.2.1)
    \item If the timer starts automatically, an option to pause, stop or hide the timer is needed (2.2.2)
    \item Exceptions include real-time events, essential time limits, 
    and times longer than the 20-hour limit (2.2.1)
\end{enumerate}


\subsubsection{Auto-Play}
One of three examples would require a fundamental change due to the guideline \emph{2.2.2 Stop, Pause, Hide}. For the analyzed example, considered fundamentally inaccessible, a video was used as a GIF image, replaying without providing an option to stop, pause, or hide the video. This may not strictly match the definition of Auto-Play in the ontology~\cite{Gray2024}, since a new video does not appear after it is finished. Still, this pattern demonstrates how such a DP would need to change. Additionally, while not found in our cohort, if Auto-Play changes the context (e.g., through redirection), there needs to be an option to turn this behaviour off, due to guideline \emph{3.2.5 Change on Request}.

For videos that use Auto-Play, the following needs to be true:
\begin{enumerate}
    \item Videos should be able to be stopped, paused, or hidden (2.2.2)
    \item An option to disable this behaviour is needed.
\end{enumerate}

\subsubsection{Hidden Information}
Two of four examples were considered fundamentally inaccessible for the same reason: a link was hidden by using the same font and background color, hindering the user from unsubscribing to a mailing list.
This is inaccessible, since the element is visually completely hidden, directly violating \emph{1.4.1 Use of Color}, \emph{1.4.3 Contrast (Minimum)}, and \emph{1.4.6 Contrast (Enhanced)}. \emph{1.4.1} requires elements to be distinguishable by color, while \emph{1.4.3} and \emph{1.4.6} require a color contrast ratio of 4.5:1+ (AA) and 7:1+ (AAA). Notably, this does not hinder use of other forms of Hidden Information, where links have the same color as the text around them. However, links must be visually distinguishable from normal text (e.g., by underlining) and fulfill the given contrast minimum ratio for WCAG compliance.

\subsubsection{Tricky but Possible Breaks in Guidelines}
Judging violations of accessibility guidelines was sometimes tricky. We were strict, but more nuance could reveal links between deception and access violations. For example, Trick Questions are often short but worded confusingly on purpose and may thus be related to \emph{3.1.5 Reading Level}, which requires simple over complicated text. The DPs Wrong Language and Complex Language may be grouped together as ``Language Inaccessibility'' (Meso-level). Wrong Language relates to \emph{3.1.1 Language of Page} and \emph{3.1.2 Language of Parts}. The guidelines only require the metadata to clearly state the language of the page/section. While not immediately obvious, language allows users to interact more easily with the website, either by skipping or using automatic translation on the relevant section(s). Complex Language, similar to Trick Questions, might be related to \emph{3.1.5 Reading Level}.

\section{Discussion}
\label{sec:discuss}
A range of accessibility issues were found associated with DPs. While statistical analysis bore no links between DP type and accessibility level, we discovered three categories of DPs that fundamentally violate WCAG definitions: Countdown Timers, Auto-Play, and Hidden Information. In short, accessibility---based on standard guidelines---can help fight DP use on the web. Since not all DP types were analyzed, more links could be identified in future. 

\subsection{Accessibility Against DPs: A New Approach with Limits}
Accessibility standards and guidelines may represent a new tool against online deception, especially when deployed alongside laws and other methods. Still, while we found DPs implicated by the WCAG, policing might prove difficult. Disuse of the three types of DPs that fundamentally violate accessibility standards may be easier to enforce.
For the rest, accessibility standards may only limit certain implementations. Legislation could then only force disuse of WCAG-conforming DP designs, rather than disuse wholesale. Yet, integration with accessibility laws, like the EEA~\cite{eea2019}, could boost the efficacy of 
the WCAG and other tools against DPs. Essentially, these laws could act as enforcer, if clear links through a broader evidential base was established. For example, with the EAA, additional sanctions and penalties could be imposed, depending on the member state~\cite{eea2019}.
While many opt for fines~\cite{bmsgpk2024},
some countries, like Ireland, issue prison sentences for severe offenses~\cite{born2025}.
Finally, autonomy may be an important part of DP legislation~\cite{Brenncke2023,Ahuja2022autonomy,Ahuja2025autonomy,santos2024online}, like transparency is for accessibility~\cite{Gkotsopoulou2023}. 
The upcoming \href{https://www.w3.org/WAI/standards-guidelines/wcag/wcag3-intro/}{WCAG 3.0}~\cite{WCAG3} could improve overall accessibility by drawing on the legal realm, perhaps by specifying transparency and autonomy as ``normative lenses.'' Future work should explore additional complementary linkages between access and law.

The strategy of accessibility enforcement---in law or standards---has understandable and clear limits. 
The same fears about effectiveness exist 
for DP laws~\cite{zdunska2024factors}, like enforcement difficulties and economic losses. However, accessibility can expose DPs and help designers avoid DP use when requested by a company, following design trends, or by accident. The availability of vetted, automated tools for accessibility compliance is also a strength. Consent management platforms (CMPs) have been criticized for use of DPs like Visual Prominence and poor accessibility~\cite{Clarke2024, Nouwens2020}, leading to research on cookie banner design~\cite{Kocyigit2023}. Similar research for objective DP classification and detection has recently increased~\cite{Kodandaram2023, Kocyigit2024, LiMeng2024}. While none focus on accessibility, associating automatically detectable WCAG guidelines with identifiable DPs could enable quick detection of deceptive \emph{and} inaccessible cases. Then, accessibility laws could be utilized, should consumer protection laws fail.

\subsection{Should Deceptive Patterns be Made Accessible? A Question of Ethics}
\label{sec:accvsinacc}

Deceptive and manipulative elements may be a core feature of certain web experiences. 
Yet, people with disabilities encounter clear barriers when interacting with DPs~\cite{Kodandaram2023,Lewis2025}. This raises the question: Should we make inaccessible DPs accessible, even in the short-term, while laws catch up? One argument against this would be to help people with disabilities dodge DPs. However, this depends on the type of DP and the severity of the outcome. For example, an inaccessible version of Sneak into Basket could lead people with disabilities to fall for the additional costs more easily. Such cases could make a lawsuit easier to pursue. Yet, people interacting with an accessible version of the DP could detect the trick and avoid paying additional costs to begin with. We note that interactions with DPs cannot be generalized for all users with disabilities. Users with ADHD, for example, may have an advantage when avoiding certain types of DPs~\cite{mildner2025adhd}. We do not argue for or against making DPs accessible, but highlight that there is no simple solution in the context of accessibility. A similar argument would be to punish providers that rely on DPs by preventing access to certain consumers groups. However, consumers should have the right to choose, even at the risk of deception. Also, some providers may be the \emph{only} providers of essential services, like replacement parts for wheelchairs. Conversely, one example from \citet{Lewis2025} features Countdown Timers, where the user was informed through audio, in intervals, about the timer's state. While this was WCAG-accessible, the user complained that it distracted from the intended task. Thus, the WCAG does not always help against DPs, with such examples demonstrating the complexities and nuances of this approach. This surfaces an underlying issue where accessibility has the potential to clash with usability, specifically for users with different disabilities. We offer no easy answers, but echo previous research~\cite{Bigham2017, Sauer2020, Lewis2025} by noting the issue, especially given how our work may aid efforts to redress DP use.

\section{Limitations and Future Work}
\label{sec:limits}
Since the WCAG only applies to web content, we had to exclude most sourced examples, e.g., screenshots and smartphone-only examples. However, we did use screenshots of inaccessible websites (currently or through the Wayback Machine) that could be analyzed with color contrast tools because the main deception happened through a visual element (\autoref{sec:procedure}). This might slightly skew results, since non-visual forms of deception could not be assessed. 
We also struggled to find a variety of DPs. 
Some varieties were more common, likely due lack of knowledge and reporting of other types~\cite{Mathur2019, Mildner2023DarkArts}. Differences in type frequency also depended on the medium (for example, Intermediate Currency in video games)~\cite{Niknejad2024}, leading to certain types appearing less often as web content.  

Future work on the intersection of the WCAG and DPs could isolate the DPs most relevant to web interaction. What specific elements are most relevant to deception should also be clearly defined. Furthermore, although accessibility may improve usability for all users~\cite{Schmutz2017, Aizpurua2016}, it can sometimes create new problems~\cite{Bigham2017, Sauer2020, Lewis2025}. 
Exploring DPs at this intersection may reveal more about this relationship. Finally, given the upcoming \href{https://www.w3.org/WAI/standards-guidelines/wcag/wcag3-intro/}{WCAG 3.0}~\cite{WCAG3}, using DPs as a benchmark to improve guidelines could also improve accessibility, potentially hindering use of DPs and leading to more efficient court cases, especially with the help of automatic DP and WCAG detection methods.

\section{Conclusion}
\label{sec:conclude}
In this exploratory study, we proposed the use 
the WCAG to combat deceptive patterns online. We analyzed 68 examples using automated tools and manual analysis via the WCAG. We found a range of accessibility issues, even with no significant difference in number of in/accessible examples or issues by DP type. However, when comparing specific examples against specific WCAG guidelines, we discovered crucial relationships where some guidelines inhibited some properties or the entire design of certain DPs. 
A more nuanced approach to interpreting the guidelines may reveal more cases. Our work is a proof-of-concept first step towards uniting access and usability against web-based deception.



%




\begin{acks}
Thank you to the members of the Aspirational Computing Lab for general support. This work was funded in part by the Institute of Science Tokyo Young Science and Engineering Researchers Program (YSEP) and a 
Japan Science and Technology Agency (JST) PRESTO grant (\#JPMJPR24I6).
\end{acks}

\bibliographystyle{ACM-Reference-Format}
\balance
\bibliography{REFS}






\end{document}